\documentclass[aps,prl,twocolumn,preprintnumbers,showpacs,nofootinbib]{revtex4}

\usepackage[english]{babel}
\usepackage{amsmath}
\usepackage{amssymb}
\usepackage{epsfig}
\usepackage{graphics,psfrag,rotating}
\usepackage{graphicx}
\usepackage{dcolumn}
\usepackage{bm}
\usepackage{float}
\begin{document}
\newcommand{\Od}{{\cal O}}
\newcommand{\lsim}   {\mathrel{\mathop{\kern 0pt \rlap
  {\raise.2ex\hbox{$<$}}}
  \lower.9ex\hbox{\kern-.190em $\sim$}}}
\newcommand{\gsim}   {\mathrel{\mathop{\kern 0pt \rlap
  {\raise.2ex\hbox{$>$}}}
  \lower.9ex\hbox{\kern-.190em $\sim$}}}


\title{Gravitational collapse in $f(R)$ theories}%

\author{J.\,A.\,R.\,Cembranos$^{(a)}$\footnote{E-mail: cembranos@physics.umn.edu},
A.\,de la Cruz-Dombriz$\,^{(b, c)}$\footnote{E-mail: alvaro.delacruzdombriz@uct.ac.za}
and B.\,Montes N\'u\~nez$\,^{(d)}$\footnote{E-mail: barbara.montes@ciemat.es}
}
\affiliation{$^{(a)}$Departamento de F\'{\i}sica
Te\'orica I, Universidad Complutense de Madrid, E-28040 Madrid,
Spain}

\affiliation{$^{(b)}$   Astrophysics, Cosmology and Gravity Centre (ACGC), University of Cape Town, Rondebosch, 7701, South Africa}

\affiliation{$^{(c)}$ Department of Mathematics and Applied Mathematics, University of Cape Town, 7701 Rondebosch, Cape Town, South Africa}

\affiliation{$^{(d)}$ 
Centro de Investigaciones Energ\'eticas, Medioambientales y Tecnol\'ogicas (CIEMAT), 28040 Madrid, Spain
 }

\date{\today}

\begin{abstract}
We study the gravitational collapse in modified gravitational theories. In particular, we analyze a general $f(R)$ model with uniformly collapsing cloud of self-gravitating dust particles. This analysis shares analogies with the formation of large-scale structures
in the early Universe and with the formation of stars in a molecular cloud experiencing gravitational collapse. In the same way, this investigation
can be used as a first approximation to the modification
that stellar objects can suffer in these modified theories of gravity.
We study concrete examples, and find that the analysis of gravitational collapse is an important tool to constrain models that present late-time cosmological acceleration.
\end{abstract}

\maketitle

\section{Introduction}
In the general study of astrophysical weak gravitational fields, relativistic effects tend to be ignored. However, there are clear
examples of stellar objects in which these effects may have important consequences, such as neutron stars, white dwarfs,
supermassive stars or black holes. Indeed, it becomes necessary to consider observationally consistent gravitational theories
to study these objects. General Relativity (GR) has been the most widely
used theory but other gravitational theories may be studied for a better understanding of the features and properties of such objects and to compare their predictions with experimental results.

The gravitational collapse for a spherically symmetric stellar object has been extensively studied in the GR framework (see \cite{Weinberg} and references therein). By assuming the metric of the space-time to be
spherically symmetric and that the collapsing fluid is pressureless, the found metric interior to the object turns to be
Robertson-Walker type with a parameter playing the role of spatial curvature and proportional to initial density. The time lapse and the size of the object are given by  a cycloid parametric equation with an angle parameter $\psi$.  Further results are that the time when the object gets zero size is finite and inversely proportional to the square root of the initial density.
Finally,  the redshift seen by an external observer is nevertheless infinite when time approaches the collapse time.

In spite of the fact that GR has been one of the most successful theories of the twentieth century, it does not give a satisfactory explanation to some of the latest cosmological and astrophysical observations with usual matter sources. In the first place, a dark energy contribution needs to be considered to provide cosmological acceleration whereas the baryonic matter content has to be supplemented by a dark matter (DM) component to give a satisfactory description of large scale structures, rotational speeds of galaxies, orbital velocities of galaxies in clusters, gravitational lensing of background objects by galaxy clusters, such as the Bullet Cluster, and the temperature distribution of hot gas in galaxies and clusters of galaxies.
All these evidences have revealed the interest to study alternative cosmological theories. This extra DM component is required to account for about $20\%$ of the energy content of our Universe. Although there are many possible origins for this component \cite{DM}, DM is usually assumed to be in the form of thermal relics that naturally freeze-out with the right abundance in many extensions of the standard model of particles \cite{WIMPs}. Future experiments will be able to discriminate among the large number of candidates and models, such as direct and indirect detection designed explicitly for their search \cite{isearches}, or even at high energy colliders, where they could be produced \cite{Coll}.

A larger number of possibilities can be found in the literature to generating the present accelerated expansion of the Universe \cite{DE}.
One of these methods consists of modifying Einstein's gravity itself \cite{otras, Tsujikawareview2010}
without invoking the presence of any exotic dark energy among the cosmological components.
In this context, functions of the scalar curvature when included in the gravitational action
give rise to the so-called $f(R)$ theories of modified gravity
\cite{Tsujikawa_Felice_fR_theories}. They amount to modifying the l.h.s. of the
corresponding equations of motion and provide a geometrical origin to the accelerated cosmological
expansion. Although such theories are able to describe
the accelerated expansion on cosmological scales correctly  \cite{delaCruzDombriz:2006fj}, they typically
give rise to strong effects on smaller scales. In any case,  viable models
can be constructed to be compatible with local gravity tests and other
cosmological constraints \cite{varia}.


The study of alternative theories of gravitation requires establishing methods able to
confirm or discard their validity by studying the cosmological
evolution,  the growing of cosmological perturbation and, at astrophysical level, the existence
of objects predicted by GR such as black holes \cite{delaCruzDombriz:2009et} 
or dust clouds forming compact structures. It is well-known that $f(R)$ gravity theories may mimic any cosmological evolution
by choosing adequate $f(R)$ models, in particular that of $\Lambda\text{CDM}$  \cite{delaCruzDombriz:2006fj}. This is the so-called {\it degeneracy problem}
that some modified gravity theories present:
accordingly, the exclusive use of observations such as high-redshift Hubble diagrams from SNIa \cite{Riess}, baryon
acoustic oscillations \cite{BAO} or CMB shift factor \cite{Spergel}, based on different distance measurements which are sensitive only to the
expansion history, cannot settle the question of the nature of dark energy \cite{Linder} since identical results may be
explained by several theories. Nevertheless, it has been proved that $f(R)$ theories - even mimicking
the standard cosmological expansion - provide different results from  $\Lambda\text{CDM}$ if the scalar cosmological
perturbations are studied \cite{delaCruzDombriz:2008cp}.
Consequently, the power spectra would be distinguishable from that predicted by $\Lambda\text{CDM}$ \cite{PRL_Dombriz}.

It is therefore of particular interest, to establish the predictions of $f(R)$ theories concerning the gravitational collapse, and in particular collapse times, for different astrophysical objects. Collapse properties
may be either exclusive for Einstein's gravity or intrinsic to any covariant gravitation theory. On the other hand, obtained results may be shed some light
about the models viability and be useful to discard models in disagreement
with expected physical results.

In \cite{Sharif:2010um} the authors studied gravitational collapse of a spherically symmetric perfect fluid in $f(R)$ gravity.
By proceeding in a similar way to \cite{Weinberg}, the object mass was deduced from the junction conditions
for interior and exterior metric tensors.  Finally, they concluded that $f(R_0)$
(constant scalar curvature term) slows down the collapse of matter and plays the role of a cosmological constant.
Authors in \cite{Bamba:2011sm} paid attention to the curvature singularity appearing in the star collapse process in $f(R)$ theories. This singularity was claimed to be generated in viable $f(R)$ gravity and can be avoided by adding a $R^{\alpha}$ term. They also studied exponential gravity and the time scale of the singularity appearance in that model. It was shown that in case of star collapse, this time scale is much shorter than the age of the universe.  Analogous studies were carried on by \cite{Arbuzova:2010iu}
claiming that in this class of theories,
explosive phenomena in a finite time may appear in systems with time dependent increasing mass density.

Reference \cite{Santos:2011ye} includes a complete study of neutron stars in  $f(R)$ theories. The most relevant result in this investigation suggests that $f(R)$ theory allows stars in equilibrium with arbitrary baryon number, no matter how large they are. Very recently authors in \cite{Hwang} studied collapse of charged black holes by using the double-null formalism.

Charged black holes in f(R) gravity can have a new type of singularity due to higher curvature corrections, the so-called f(R)-induced singularity, although it is highly model-dependent.

The present work has been arranged as follows: in section II, $f(R)$ modified gravity theories will be introduced. Gravitational collapse in $f(R)$ theories will be presented in section III. After performing some calculations, the evolution equation for the object scale factor will be obtained. This equation will be used throughout the following sections.  Section IV is then dedicated to achieve solutions for the modified equations in three qualitatively different $f(R)$ models, which try to illustrate the broad phenomenology of the subject. This is therefore the aim of this section: to study gravitational collapse by calculating the evolution of the object scale factor in particular $f(R)$ models. Finally, the conclusions based upon the presented results will be analyzed in detail in section V.

\section{II. $f(R)$ theories of gravity}

With the aim of proposing and alternative theory to GR, a possible modification consists of adding a function of the scalar curvature, $f(R)$,
to the Einstein-Hilbert (EH) Lagrangian. Therefore the gravitational action becomes
\footnote{In the present work we employ the natural units system in which $\hbar=c=1$.
Note also that our definition for the Riemann tensor is
    $R_{\mu\nu\kappa}^{\sigma}=\partial_{\kappa}\Gamma_{\mu\nu}^{\sigma}-\partial_{\nu}\Gamma_{\mu\kappa}^{\sigma}+
    \Gamma_{\kappa\lambda}^{\sigma}\Gamma_{\mu\nu}^{\lambda}-\Gamma_{\nu\lambda}^{\sigma}\Gamma_{\mu\kappa}^{\lambda}$.}
\begin{equation}
S_G=\frac{1}{16\pi G}\int \text{d}^4x \sqrt{\vert g \vert}\left(R+f(R)\right).
\label{Modified action}
\end{equation}
By performing variations with respect to the metric, the modified Einstein equations turn out to be
\begin{equation}
(1+f_R)R_{\mu\nu}-\frac{1}{2}(R+f(R))g_{\mu\nu}+{\cal D}_{\mu\nu}f_R\,=\,-8\pi G \,T_{\mu\nu},
\label{fieldtensorialequation}
\end{equation}
where $T_{\mu\nu}$ is the energy-momentum tensor of the matter content, $f_R\,\equiv\,\text{d}f(R)/\text{d}R$ and
${\cal D}_{\mu\nu}\equiv \nabla_{\mu}\nabla_{\nu}-g_{\mu\nu}\square$ with $\square\,\equiv\,\nabla_{\alpha}\nabla^{\alpha}$ and $\nabla$ is the usual
covariant derivative.

These equations may be written \textit{\`{a} la Einstein} by isolating on the l.h.s. the Einstein tensor and the $f(R)$ contribution on the r.h.s. as follows
\begin{eqnarray}
R_{\mu\nu}&-&\frac{1}{2}R g_{\mu\nu}
\,=\,\frac{1}{(1+f_R)}\left[\vphantom{\frac{1}{2}}-8\pi G T_{\mu\nu}
\right. \nonumber\\ &-& \left.
{\cal D}_{\mu\nu}f_R+\frac{1}{2}\left(f(R)-Rf_R\right)g_{\mu\nu}\right]
\end{eqnarray}
We can also find the expression for the scalar curvature by contracting \eqref{fieldtensorialequation} with $g^{\mu\nu}$ which gives:
\begin{eqnarray}
(1-f_R)R+2f(R)+3\square f_R\,=8\pi G \,T.
\label{}
\end{eqnarray}
Note that, unlike GR where $R$ and $T$ are related algebraically, for a general $f(R)$ those two quantities are dynamically related.
In the homogeneous and isotropic case, the scalar curvature in $f(R)$ theories becomes
\begin{eqnarray}
R=\frac{8\pi G \,T-2f(R)-3 \ddot{f}_R\,}{(1-f_R)}
\label{Scalar curvature in f(R)}
\end{eqnarray}
where dot means the derivative with respect to cosmic time.

\section{III. Gravitational collapse in $f(R)$}
In the case of our investigation, we introduce the spherically symmetric metric   
\begin{equation}
\text{d}s^2 =\text{d}t^2-U(r,t)\text{d}r^2-V(r,t)(\text{d}\theta^2+\text{sin}^2\theta \text{d}\phi^2)
\label{Intervalo colapso}
\end{equation}
If the collapsing object is approximated to be pressureless $p\simeq 0$, the components of the energy-momentum tensor can be expressed as follows
\begin{eqnarray}
T_{\mu\nu}=\rho u_{\mu}u_{\nu}\;;\;
T^{t}_{\;\;t}=\rho\;;\;
T^{i}_{\;\;i}=0 \;\; \text{if}\;\; i=r,\theta,\,\phi.
\label{pressureless_fluid}
\end{eqnarray}
We may further simplify the collapse model by considering
$\rho$ independent from the position. Therefore, we can
search -as is actually the usual approach in the GR case-  a separable solution
for this metric as follows
\begin{equation}
    U(r,t)=A_1^2(t)h(r),\quad V(r,t)=A_2^2(t)r^2,
\label{Funciones U y V en f(R)}
\end{equation}
where a previous reparametrization of the radial coordinate is required.
When $f(R)$ modified tensorial equations are studied in the homogeneous and isotropic case - in which
$f(R)$ does not depend on the position-, the trace component provides
\begin{eqnarray}
\left(\frac{\dot{A}_2}{A_2}-\frac{\dot{A}_1}{A_1}\right)\frac{g'}{g}=0\Rightarrow \frac{\dot{A}_2}{A_2}=\frac{\dot{A}_1}{A_1}.
\label{A1_vs_A2}
\end{eqnarray}
From \eqref{A1_vs_A2}, we deduce that $A_1$ and $A_2$ are proportional, in other words, $A_1(t)=C(r)A_2(t)$. So, if
we choose $A_1(t)=A_2(t)\equiv A(t)$, the dependence in the radial coordinate is reabsorbed by $h(r)$. Hence:
\begin{equation}
    U(r,t)=A^2(t)h(r),\quad V(r,t)=A^2(t)r^2.
\label{U and V general functions}
\end{equation}
%
Components $tt$, $rr$ and $\theta\theta$ for the modified tensorial equations may be written respectively in terms of the functions $A(t)$ and $h(r)$ as follows
\begin{eqnarray}
&&3\frac{\ddot{A}}{A}=\frac{1}{(1+f_R)}\left[-8\pi G\rho+
3\frac{\dot{A}}{A}\dot{f}_{R}+
\frac{1}{2}\left(R+f(R)\right)\right],\nonumber\\
&&
\label{Rtt with U and V simplified}
\end{eqnarray}
\begin{eqnarray}
&&A\ddot{A}+2\dot{A}^2+\frac{h'}{r h^2}
=\frac{A^2}{(1+f_R)}\left[
\ddot{f}_{R}\,
+2\frac{\dot{A}}{A}\dot{f}_{R}
\right. \nonumber\\ && \left.
+\frac{1}{2}\left(R+f(R)\right)\right],
\label{Rrr with U and V simplified}
\end{eqnarray}
\begin{eqnarray}
&&A\ddot{A}+2\dot{A}^2+\frac{1}{r^2}-\frac{1}{hr^2}
+\frac{h'}{2r h^2}\nonumber\\
&=&\frac{A^2}{(1+f_R)}\left[
\ddot{f}_{R}
+2\frac{\dot{A}}{A}\dot{f}_{R}
+\frac{1}{2}\,(R+f(R))\right].\nonumber\\
\label{Rthth with U and V simplified}
\end{eqnarray}
%
Let us point out two important aspects of equations \eqref{Rrr with U and V simplified} and \eqref{Rthth with U and V simplified}: firstly, terms on the r.h.s of both equations are equal. Secondly, the term on the l.h.s. exclusively depends on $r$, whereas the term on the r.h.s. only depends on $t$ in both equations, so that they must be constants\footnote{This fact is also satisfied in the GR case and allows the simplification of the calculus.}. Therefore we may equal l.h.s. of both equations to provide
\begin{eqnarray}
\frac{1}{r}\frac{h'}{h^2}=\frac{1}{r^2}-\frac{1}{hr^2}+\frac{1}{2r}\frac{h'}{h^2}\equiv 2k\,,
\label{Eq para constante k in modified case}
\end{eqnarray}
where we have equaled both equations (multiplied by a factor $A^2$) to a 
constant $-2k$. The resulting solution
is $h(r)=(1-kr^2)^{-1}$.
Once we have calculated $h(r)$, the resulting metric 
can be expressed as follows:
\begin{equation}
\text{d}s^2 =\text{d}t^2-A^2(t)\left[\frac{\text{d}r^2}{1-kr^2}+r^2(\text{d}\theta^2+\text{sin}^2\theta \text{d}\phi^2)\right]\,,
\label{M�trica homog�nea y is�tropa in modified case}
\end{equation}
which is formally the same as the one obtained in the GR case \cite{Weinberg}. 
Expression \eqref{Eq para constante k in modified case} for $k$ may be substituted in either expression \eqref{Rrr with U and V simplified} or \eqref{Rthth with U and V simplified} yielding:
\begin{eqnarray}
&-&\frac{\ddot{A}}{A}-2\left(\frac{\dot{A}}{A}\right)^2-\frac{2k}{A^2}=\frac{1}{(1+f_R)}\left[
-\ddot{f}_{R}
-2\frac{\dot{A}}{A}\dot{f}_{R}
\right. \nonumber\\ &-& \left.
\frac{1}{2}\left(R+f(R)\right)\right].
\label{eq casi casi}
\end{eqnarray}
Taking into account $\rho\,(t)=\rho\,(t=0)/A(t)^3$ (given by the energy motion equation for dust matter) and the results in (\ref{eq casi casi}), 
equation \eqref{Rtt with U and V simplified} becomes:
\begin{eqnarray}
\dot{A}^2&=&-k+\frac{1}{(1+f_R)}\left[\frac{4}{3}\pi G \rho(0)A^{-1}
+ \frac{1}{2}A^2\ddot{f}_{R}
\right. \nonumber\\ &+& \left.
\frac{1}{2}A\dot{A}\dot{f}_{R}
+\frac{A^2}{6}\left(R+f(R)\right)\right].
\label{Eq for dotA}
\end{eqnarray}
Furthermore, provided that the fluid is assumed to be at rest for $t=0$, initial conditions $\dot{A}(t=0)=0$ and $A(t=0)=1$ hold. This last condition means that the scale factor of the object at initial time is normalized to unity. In order to simplify the notation we define $R(t=0)\equiv R_0$ and $\rho\,(t=0)\equiv \rho_0$. Therefore, evaluation of \eqref{Eq for dotA} at $t=0$ allows to recast $k$ as follows
\begin{eqnarray}
k&=&\frac{1}{(1+f_R(R_0))}\left[\frac{4\pi G}{3}\rho_0
+\frac{1}{2}\ddot{f}_{R}(R_0)
+\frac{1}{6}(R_0+f(R_0))\right]
\nonumber\\
&&\label{Curvature in terms of R(0)}
\end{eqnarray}
Once $k$ has been expressed in terms of different quantities initial values, 
equations (\ref{Scalar curvature in f(R)}) and (\ref{Curvature in terms of R(0)}) may be inserted in (\ref{Eq for dotA}) to provide 
\begin{widetext}
\begin{eqnarray}
\dot{A}^2&=&-\frac{1}{6(1-f_R^2(R_0))}\left[\vphantom{\ddot{f}_R(R_0)}\,8\pi G\rho_0\left(2-f_R(R_0)\right)
- f(R_0)(1+f_R(R_0))-3\ddot{f}_R(R_0)f_R(R_0)\,\right]
+\frac{1}{(1-f_R^2)}\,\frac{8\pi G}{3}\rho_0\,A^{-1}\nonumber\\
&&-\frac{1}{6(1-f_R^2)}\left[8\pi G\rho_0\,A^{-1}f_R+3A^2\ddot{f}_{R}f_R-
3A\dot{A}\dot{f}_{R}(1-f_R)+A^2f(R)(1+f_R)\right].
\label{Simplified Eq for dotA}
\end{eqnarray}
\end{widetext}
The previous expression will be solved perturbatively to first order in perturbations for different $f(R)$ models.
Let us remind at this stage that the zeroth order solution of GR is given by the parametric equations of a cycloid \cite{Weinberg}:
\begin{equation}
t=\frac{\psi+\text{sin}\,\psi}{2\sqrt{k}},\;\;\; A_G=\frac{1}{2}(1+\text{cos}\,\psi).
\label{Eqs cicloide}
\end{equation}
Expression \eqref{Eqs cicloide} clearly implies that a sphere with initial density $\rho_0$ and negligible  pressure will collapse from rest to a state of infinite proper energy density in a finite time that we will denote $T_G$. This time is obtained for the first value of $\psi$ such as $A_G=0$, i.e. for $\psi=\pi$. It means
\begin{eqnarray}
T_G=\left(\frac{\pi+\text{sin}\pi}{2\sqrt{k}}\right)=\frac{\pi}{2\sqrt{k}}=\frac{\pi}{2}\left(\frac{3}{8\pi G\rho_0}\right)^{1/2}.
\label{Valor de t para colapso}
\end{eqnarray}

In order to study the modification to the gravitational collapse in $f(R)$ theories, we will expand $A$ around $A_G$ and $f(R)$ around the
scalar curvature in GR ($R=R_G$):
\begin{equation}
A=A_G+g(\psi)\,,
\label{Series expansion for A}
\end{equation}
\begin{equation}
f(R)\simeq f(R_G)+f'(R_G)(R-R_G).
\label{Series expansion for f(R)}
\end{equation}
The presence of a function $f(R)$ in the gravitational Lagrangian will represent a
correction of first order 
with respect to the usual EH Lagrangian. Hence, $g(\psi)$ as defined in (\ref{Series expansion for A})
will be also first order at least.
%
By substituting the series expansions \eqref{Series expansion for A} and \eqref{Series expansion for f(R)} 
in expression \eqref{Simplified Eq for dotA} until first order in $\varepsilon$,  we find
that equation (\ref{Simplified Eq for dotA}) becomes
\begin{widetext}
\begin{eqnarray}
\text{tg}\left(\frac{\psi}{2}\right)g'\,&=&\,-\frac{1}{2}\text{cos}^{-2}\left(\frac{\psi}{2}\right) g
+\frac{1}{12k}\text{cos}^{2}\left(\frac{\psi}{2}\right)\left(f(R_{G0})+3 k f_{R}(R_{G0})\right)
-\frac{1}{4}f_R(R_G)\nonumber\\
&&+\frac{1}{4\sqrt{k}}\,\text{sin}^{}\left(\frac{\psi}{2}\right)\text{cos}^{3}\left(\frac{\psi}{2}\right)\dot{f}_{R}(R_G)
-\frac{1}{12k}\,\text{cos}^{6}\left(\frac{\psi}{2}\right) f(R_G)\,,
\label{Eq g' with generic Rs}
\end{eqnarray}
\end{widetext}
%
where we have cancelled out the GR exact solution and only kept first order perturbed terms.
Equation \eqref{Eq g' with generic Rs} will provide $g(\psi)$ evolution 
for different $f(R)$ models to be considered in the next section.

\section{IV. $f(R)$ theory results}
In this section we shall consider three illustrative $f(R)$ models and study the gravitational
collapse process for collapsing dust. The models under consideration are
\subsection{Model 1: $f(R)=\varepsilon R^2$}
This function has been proposed both as a viable inflation candidate
\cite{Starobinsky:1980te} and as a dark matter model \cite{R2DM}. In this last reference, the $\varepsilon$ parameter definition reads
\begin{eqnarray}
\varepsilon=\frac{1}{6m_0^2}\, ,
\label{Eq epsilon1}
\end{eqnarray}
and the minimum value allowed for $m_0$ is computed as
$m_0=2.7\times10^{-12}$ GeV at 95 $\%$ confidence level, i.e. $\varepsilon\leq 2.3\times10^{22}\,\text{GeV}^{-2}$. On the
other hand, $\varepsilon>0$ is needed to ensure the stability of the model, since in the opposite case, a tachyon is present in the theory. These constraints are in agreement with \cite{Berry}.

\begin{figure}
\begin{center}
\resizebox{8.5cm}{8.5cm}
{\includegraphics{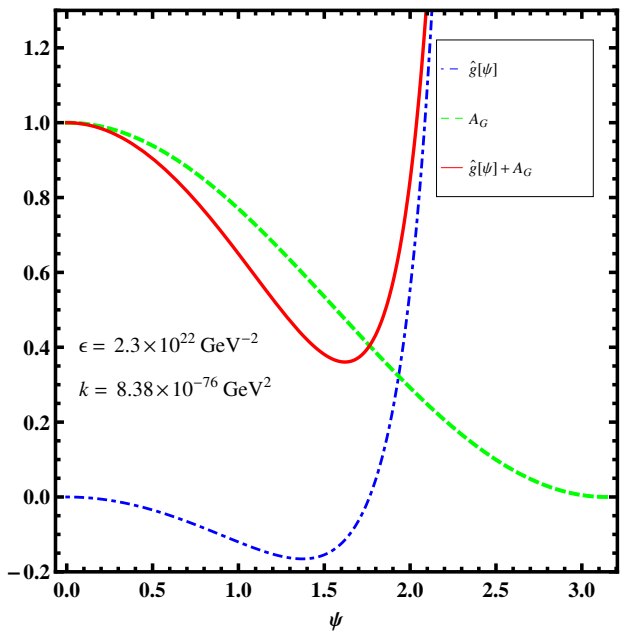}}
\caption {\footnotesize
Comparison between the solution given by \eqref{Analytical solution for g model 1} setting $c_1=0$ and the GR case for the model 1: $f(R)= \varepsilon R^2$. The plotted $k$ value is fixed by Eq. (\ref{Curvature}), where the density is $\rho_{\text{SF}}\simeq1.5 \times 10^{-38}$ GeV$^4 \simeq 3.5\times10^{-18}\,{\text{kg}}/{\text{m}^3}$, i.e., the matter density in the early Universe at redshift $z\simeq 1100$
marking the decoupling of matter and radiation and the beginning of structure formation (SF). The modification is extraordinarily small and
has been increased 52 orders of magnitude to make it observable: $\hat{g}(\psi)=10^{52} g(\psi)$.}
\label{Figure_Intersection_1}
\end{center}
\end{figure}

After some algebra, for this model, equation (\ref{Eq g' with generic Rs})
can be written as follows:
\begin{eqnarray}
&&g'(\psi )+g(\psi ) \csc (\psi )+\frac{9}{8}k\varepsilon\left(\sin (\psi )+2 \tan
\left(\frac{\psi }{2}\right)
\right. \nonumber\\ &-& \left.
4 \tan ^4\left(\frac{\psi }{2}\right)
\csc (\psi )\right)=0.
\label{Simplified Final eq for g' model 1}
\end{eqnarray}
The homogeneous equation associated to (\ref{Simplified Final eq for g' model 1}) presents the solution $g_{hom}(\psi )\propto\cot\left(\psi/2\right)$, which
diverges at $\psi=0$. Therefore, its contribution will be ignored in the upcoming analysis. The analytical full solution of \eqref{Simplified Final eq for g' model 1} becomes
\begin{widetext}
\begin{eqnarray}
&&g(\psi )= c_1 \cot \left(\frac{\psi }{2}\right)-\frac{9}{128}k\varepsilon \cot
\left(\frac{\psi }{2}\right) \left\{\vphantom{\frac{\psi}{2}}
-
16\left(\psi+ \sin (\psi)\right)
+\frac{64}{5} \tan \left(\frac{\psi }{2}\right)\left[4-\sec ^2\left(\frac{\psi}{2}\right)\left(\sec^2\left(\frac{\psi}{2}\right) - 2\right)\right]
\right\}\,,\nonumber\\
&&
\label{Analytical solution for g model 1}
\end{eqnarray}
\end{widetext}
This analytical solution given by \eqref{Analytical solution for g model 1} can be compared with the GR one by plotting them together as shown in Figure \ref{Figure_Intersection_1}.
As we see in this Figure, in the first stage of the collapse, the correction is negative, what implies that we have a larger contraction. On the contrary, very close to $\psi=\pi$, where the solution can be approximated as
\begin{eqnarray}
g(\psi )\simeq\frac{72k\varepsilon}{5(\psi-\pi)^4}.
\label{Series expansion of analytical solution for g model 1}
\end{eqnarray}
the sign of the modification changes and the total collapse is avoided. Exactly at this moment, the perturbation leaves the linear regime and a more complete analysis is required. It is interesting to estimate when the linear approach fails and an important modification is expected.

\begin{figure}
\begin{center}
\resizebox{8.5cm}{8.5cm}
{\includegraphics{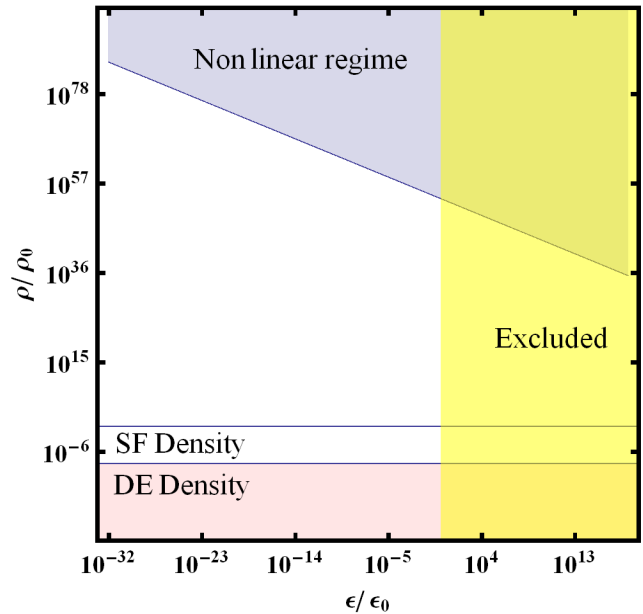}}
\caption {\footnotesize
Validity of the perturbative regime for model 1, showing different relevant regions: In blue we show the region where our linear approach loses its validity. The excluded region is depicted in yellow and determined by the condition $\varepsilon\leq 2.3\times10^{22}\,\text{GeV}^{-2}$. Finally, the density marking the beginning of structure formation (SF) and the dark energy (DE) density ($\rho_{\text{DE}}\simeq2.8 \times 10^{-47}$ GeV$^4$) have also been plotted
for reference.}
\label{Validity regime 1}
\end{center}
\end{figure}

By using the collapse time parametrization (\ref{Eqs cicloide}) in the  $\psi\rightarrow \pi$ limit, one gets
\begin{eqnarray}
t&=&\frac{\psi+\sin(\psi)}{2\sqrt{k}}\simeq \frac{\pi+1/6(\Delta\psi)^3}{2\sqrt{k}},
\label{Collapsing time series expansion model 1}
\end{eqnarray}
or written in terms of the relative variation:
\begin{eqnarray}
\frac{\Delta t}{t_{GR}}&\simeq& \frac{(\Delta\psi)^3}{6\pi}\, .
\label{Collapsing time series expansion model 1 variation}
\end{eqnarray}

We are interested in estimating the region of the parameter space of the model, where
the modified collapse, and the result for $\psi_{C}$ ($\psi$ value for the collapse) is significantly different from
the one predicted by GR ($\psi_{C,GR}\equiv \pi$). With this purpose, we can estimate the values for which $A_G$
is of the same order of its correction. As one can see in Figure \ref{Figure_Intersection_1},
this deviation is more important close to the final stage of the collapse, for $\psi \sim \pi$.
In this region, $A_G$ can be approximated by:
\begin{eqnarray}
A_G=\frac{1}{2}\left(1+\text{cos}\, \psi \right)\simeq\frac{\left(\psi-\pi\right)^2}{4}\,.
\label{A_G series expansion}
\end{eqnarray}
We can use these approximations in the limit $\psi\rightarrow\pi$ to determine the intersection between the particular solution of \eqref{Simplified Final eq for g' model 1} and the GR solution: $A_G$. This calculation will help establishing the validity regime of the perturbative approach. Therefore, by imposing $|g(\psi)|=|A_G|$,
with $g(\psi)$ given by equation (\ref{Series expansion of analytical solution for g model 1}), we obtain:
\begin{eqnarray}
\psi=\pi-\left(\frac{288|k\varepsilon|}{5}\right)^{1/6}\,.
\label{Regime validity psi value}
\end{eqnarray}

At this point, it is necessary to clarify the physical value for $k$ in order to discuss if the departure from linearity is important. $k$ is the initial condition given
by equation \eqref{Curvature in terms of R(0)} that depends on the matter density, the initial curvature and the particular $f(R)$ model. In our analysis we are interested in studying
the modification to the gravitational collapse in GR and for this reason we will assume the same value of $k$ than as given in GR. This implies
that the entire modification has a dynamical origin and it does not come from a change in the initial conditions. Therefore, we will assume that $k$ only depends on the matter density:
\begin{eqnarray}
k=\frac{8\pi G}{3}\rho_0.
\label{Curvature}
\end{eqnarray}

For the most physically interesting values of $k$ and $\varepsilon$, for which we have studied gravitational collapse of a dust matter cloud, the value of $\psi$ is quite close to $\psi=\pi$ and therefore the asymptotic approach to obtain (\ref{Regime validity psi value}) is fully justified. The results are summarized in Figure \ref{Validity regime 1},
were the non-linear regime is shown for different values of $\epsilon$ and initial densities.
For example, it is interesting to check the behavior for the matter density in the
early Universe at redshift $z\simeq 1100$, which marks the decoupling of matter and radiation and the beginning of structure formation (SF):
$\rho_{\text{SF}}\simeq1.5 \times 10^{-38}$ GeV$^4 \simeq 3.5\times10^{-18}\,{\text{kg}}/{\text{m}^3}$. In this particular case, the calculated root
presents a slight difference with respect to the solution for GR. In particular, $\psi_{C}$ differs from $\psi_{C,\,\text{GR}}\equiv \pi$ in the ninth significant figure.

Although this modification is not detectable for this first model as we deduce from the last considerations, it is interesting to stress
that the relative modification is higher for denser media since the correction increases with  $\rho_0$ as  $\Delta t/t_{GR} \propto \sqrt{\rho_0}$. This
behavior is significantly different with respect to other models as we will see in the following sections. In Figure \ref{Validity regime 1}, we have represented some relevant density values as well as the non linear regime region. The loss of the linear regime takes place at high densities since the correction is directly proportional to $\rho$.

\subsection{Model 2: $f(R)=\varepsilon R^{-1}$}
We will continue our analysis with the $f(R)$ model proposed in reference \cite{Carroll:2003wy} as a dark energy candidate. This possibility is
currently excluded, but this model is a simple example that help to understand the gravitational collapse modifications in models that provide
late-time acceleration.

For this model, equation \eqref{Eq g' with generic Rs} becomes:
\begin{eqnarray}
g'(\psi )+g(\psi ) \csc (\psi )=\frac{\varepsilon}{6k^2}\sin \left(\frac{\psi }{2}\right) \cos ^{13}\left(\frac{\psi }{2}\right)\,,
\label{Simplified Final eq for g' model 2}
\end{eqnarray}
whose full solution is
\begin{widetext}
\begin{eqnarray}
g(\psi )&=& c_1 \cot \left(\frac{\psi }{2}\right)+\frac{1}{6} \frac{\varepsilon}{k^2}\left(\frac{33 \psi }{2048}+\frac{165 \sin (\psi)}{8192}
-
\frac{11 \sin (2 \psi )}{8192}-\frac{121\sin (3 \psi )}{24576}-\frac{25 \sin (4 \psi )}{8192}
\right. \nonumber\\ &-& \left.
\frac{43 \sin (5 \psi )}{40960}-\frac{5 \sin (6 \psi )}{24576}-\frac{\sin (7 \psi )}{57344}\right) \cot
\left(\frac{\psi }{2}\right).\nonumber\\
\label{Analitycal solution for g model 2}
\end{eqnarray}
\end{widetext}

\begin{figure}
\begin{center}
\resizebox{8.5cm}{8.5cm}
{\includegraphics{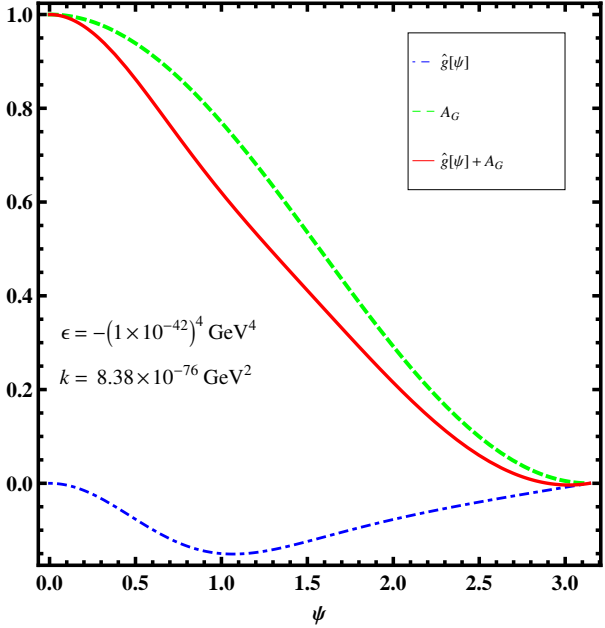}}
\caption {\footnotesize
Analogous representation to the one shown in Fig. \ref{Figure_Intersection_1}, which includes the GR solution, the modification given by \eqref{Analitycal solution for g model 2} and the sum of the two. In this figure $\hat{g}(\psi)=10^{19} g(\psi)$ in order to make the modification observable.}
\label{Figure_Intersection_2}
\end{center}
\end{figure}

In Figure \ref{Figure_Intersection_2}, it is possible to see the behavior of the modification for $\varepsilon=-\mu^4$, and $\mu=10^{-42}$ GeV as it was the value originally proposed in reference \cite{Carroll:2003wy}.

The series expansion of \eqref{Analitycal solution for g model 2} around $\psi=\pi$ reads, in this case:
\begin{eqnarray}
g(\psi )\simeq-\frac{11\pi\varepsilon(\psi-\pi)}{8192k^2}\;.
\label{Series expansion of analytical solution for g model 2}
\end{eqnarray}
Once again the intersection between the particular solution of \eqref{Simplified Final eq for g' model 2} and the GR solution $A_G$ can be determined in the $\psi\rightarrow \pi$ limit, with help of equation (\ref{A_G series expansion}). $|g(\psi)|=|A_G|$ implies
\begin{eqnarray}
\psi\simeq\pi-\frac{11\pi}{2048}\frac{|\varepsilon|}{k^2}.
\label{Regime validity psi value model 2}
\end{eqnarray}

\begin{figure}
\begin{center}
\resizebox{8.5cm}{8.5cm}
{\includegraphics{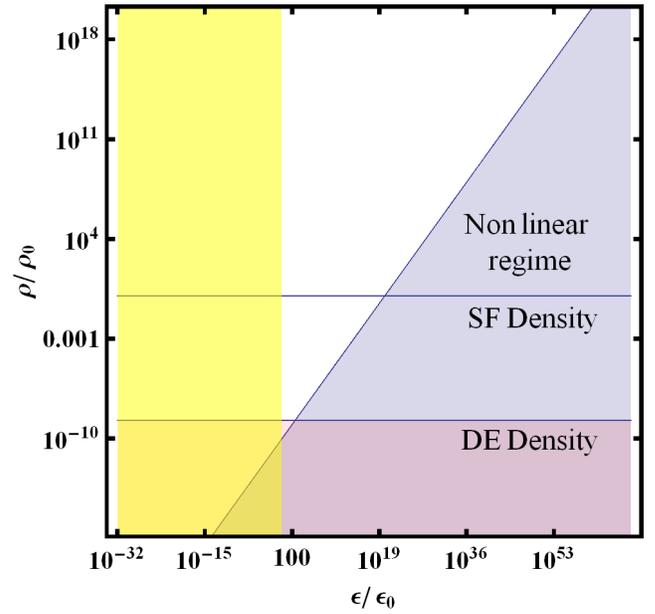}}
\caption {\footnotesize
Analogous to Figure \ref{Validity regime 1} for  model 2. The limit of the region depicted in yellow shows the proposed value
$\varepsilon=-\mu^4$, and $\mu=10^{-42}$ GeV  according to \cite{Carroll:2003wy}.}
\label{Validity model 2}
\end{center}
\end{figure}

As it can be seen in Figure \ref{Figure_Intersection_2}, the difference between the modified $\psi_{C}$ and $\psi_{C,\,\text{GR}}$ is not distinguishable for $\varepsilon=-\mu^4$ if density is higher than the standard
dark energy density. The same result is found for $\varepsilon>-\mu^4$ ($|\varepsilon|<\mu^4$ and negative). The situation changes for $\varepsilon<-\mu^4$
($|\varepsilon|>\mu^4$ and positive). This behavior can be observed in Figure \ref{Validity model 2}, where the validity of the linear regime is shown to decrease
for higher values of $|\varepsilon|$ and lower densities. As we will see in the following example, this is a general property of $f(R)$ models that provide accelerated cosmologies, at least, for densities higher than the vacuum energy.
Results in Figure \ref{Validity model 2} can be understood by estimating the correction of the
collapsing time in the linear regime as it is determined by equation (\ref{Collapsing time series expansion model 1 variation}). This $f(R)$
model provides a relative difference for the collapsing time in GR value given by
\begin{eqnarray}
\frac{\Delta t}{t_{GR}}&\simeq&-
\frac{11^3 \varepsilon^3 \pi^2}{3  k^6 2^{34}}\,.
\label{Modification time collapse model 2}
\end{eqnarray}
We observe that the correction is more negligible for denser objects. This unexpected fact can be understood since GR modification
to the scale factor is proportional to $g \propto \varepsilon/k^2$ whereas $k\propto\rho_0$. According to this dependence, a stellar
object with a higher density will suffer a less important modification and vice versa. The relative time modification is lower for denser
media since the correction decreases with  $\rho_0$ as $\Delta t/t_{GR} \propto \rho_0^{-6}$.

\subsection{Model 3: $f(R)=\lambda R_{0}\left[\left(1+\frac{R^2}{R_{0}^2}\right)^{-n}-1\right]$}
The last $f(R)$ model to be analyzed in the present work is the
well-known Starobinsky model proposed in Reference \cite{Starobinsky:2007hu}.
For this model, $n,\lambda>0$ and $R_{0}$  is considered to be of the order of the presently observed effective
cosmological constant\footnote{It is important to remark that $R_{0}$ in this context is a parameter of the model and not
the initial scalar curvature.}. With such parameter choice, this model is a viable dark energy candidate. The relation between
$\lambda$ and $R_{0}$ in vacuum is given by $H_0^2=\lambda R_0/6$ according to \cite{Starobinsky:2007hu}, where
$H_0$ is the present Hubble parameter (see \cite{WMAP} for recent WMAP data) and the
proposed value for $\lambda=0.69$.

\begin{figure}
\begin{center}
\resizebox{8.5cm}{8.5cm}
{\includegraphics{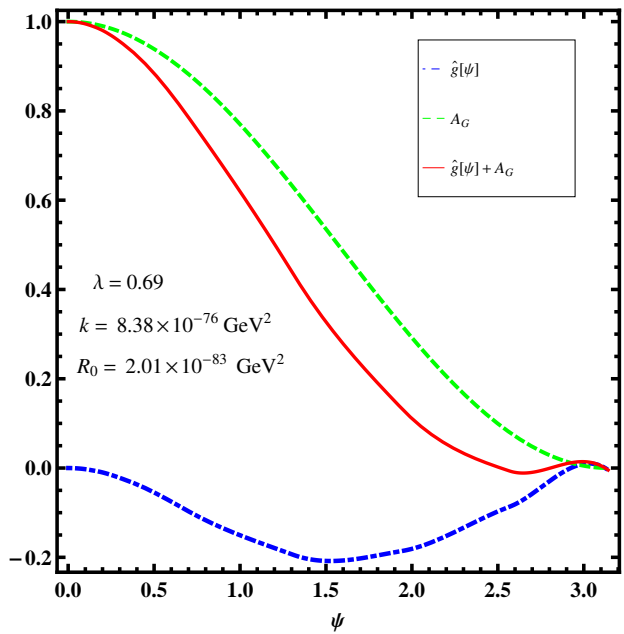}}
\caption {\footnotesize
The plotted lines are analogous to the ones in Fig. \ref{Figure_Intersection_1} and Fig. \ref{Figure_Intersection_2} but with the solution given by \eqref{Final eq for g' model 3}. In this figure $\hat{g}(\psi)=10^{9} g(\psi)$
in order to make the modification observable.}
\label{Figure_Intersection_3}
\end{center}
\end{figure}

For the sake of simplicity, let us choose $n=1$. In this case, the equation (\ref{Eq g' with generic Rs}) may be rewritten as follows:
\begin{widetext}
\begin{eqnarray}
&&-g'(\psi )-g(\psi ) \csc (\psi )-\frac{9k\lambda R_{0} \sin ^{3}(\psi)\csc^{4}\left(\frac{\psi}{2}\right)}{32\left(9k^2+R_0^2\right)^2}\left(R_0^2+3k^2\right)
+\frac{72k\lambda R_0 \sin^4\left(\frac{\psi}{2}\right)\csc(\psi)}{\left(R_0^2+9k^2\sec^{12}\left(\frac{\psi}{2}\right) \right)^2}\left(R_0^2+3k^2\sec^{12}\left(\frac{\psi}{2}\right) \right)
\nonumber\\
&&-\frac{9 k \lambda R_0^3\tan \left(\frac{\psi }{2}\right) \sec ^4\left(\frac{\psi}{2}\right) \left(R_0^2-27 k^2 \sec^{12}\left(\frac{\psi }{2}\right)\right)}{2 \left(9 k^2
\sec ^{12}\left(\frac{\psi}{2}\right)+R_0^2\right)^3}
=0.
\label{Final eq for g' model 3}
\end{eqnarray}
\end{widetext}
Unlike the other two cases, we are not able to find analytical solution for equation \eqref{Final eq for g' model 3}. Thus, specific values for  $k$ and $\lambda$ parameters  and $R_{0}$ are required to find a
numerical solution. This solution is plotted in Figure \ref{Figure_Intersection_3}.
%
%
%
In any case, equation (\ref{Final eq for g' model 3}) can be studied in the asymptotic limits
$\psi\rightarrow0$ and $\psi\rightarrow\pi$. Thus,
the corresponding series expansion of (\ref{Final eq for g' model 3}) in the $\psi\rightarrow0$ becomes
\begin{eqnarray}
&-&g'(\psi )-\frac{g(\psi )}{\psi }
-\frac{9 k \lambda R_0 \left(3 k^2-R_0^2 \right)\psi}{4 \left(9 k^2
+R_0^2\right)^2}=0\,,\nonumber\\
\label{Final Eq Psi 0 series expansion}
\end{eqnarray}
whose analytical solution is
\begin{eqnarray}
g(\psi )&=& \frac{c_1}{\psi }+\frac{3 k \lambda R_0 \left(R_0^2-3 k^2 \right)\psi ^2}{4 \left(9 k^2+R_0^2\right)^2}\,.
\label{Solution Psi 0 series expansion}
\end{eqnarray}
Since the homogeneous equation does not depend on the $f(R)$ model, the condition $c_1=0$ is also necessary in order to have a finite solution.
When the considered asymptotic limit is $\psi\rightarrow\pi$,
equation (\ref{Final eq for g' model 3}) approximately becomes
\begin{eqnarray}
-g'(\psi )+\frac{g(\psi )}{\psi-\pi }+\frac{9k \lambda R_0   (\psi-\pi)^3 \left(3 k^2+ R_0^2\right)}{32 \left(9 k^2+ R_0^2\right)^2}=0\,,\nonumber\\
\label{Eq Psi pi dominant terms simplified}
\end{eqnarray}
whose
analytical solution is
\begin{eqnarray}
g(\psi )= c_1 (\psi -\pi )+\frac{3k \lambda R_0  (\psi-\pi)^4
   \left(3 k^2+2 R_0^2\right)}{32 \left(9 k^2+ R_0^2\right)^2}.\nonumber\\
\label{Solution Psi pi series expansion}
\end{eqnarray}
This asymptotic limit of the linear correction depends on a higher power of $(\psi-\pi)$ than the GR solution  given by eq. (\ref{A_G series expansion}).
This fact implies that we cannot estimate the validity of the linear regime by using
the $\psi\rightarrow\pi$ as in the previous cases. The modification is more
important at intermediates values of $\psi$, as it can be observed in Figure
\ref{Figure_Intersection_3}. The numerical results are showed in Figure 6.
 In a similar way to the second model, for $\lambda<\lambda_0$ the modification
of the collapse is always linear and not important. The situation is different for
$\lambda>\lambda_0$, where the collapse is severely modified at densities closer
to the vacuum one. We have checked numerically
that denser environments are less affected by this gravitational model.

\begin{figure}
\begin{center}
\resizebox{8.5cm}{8.5cm}
{\includegraphics{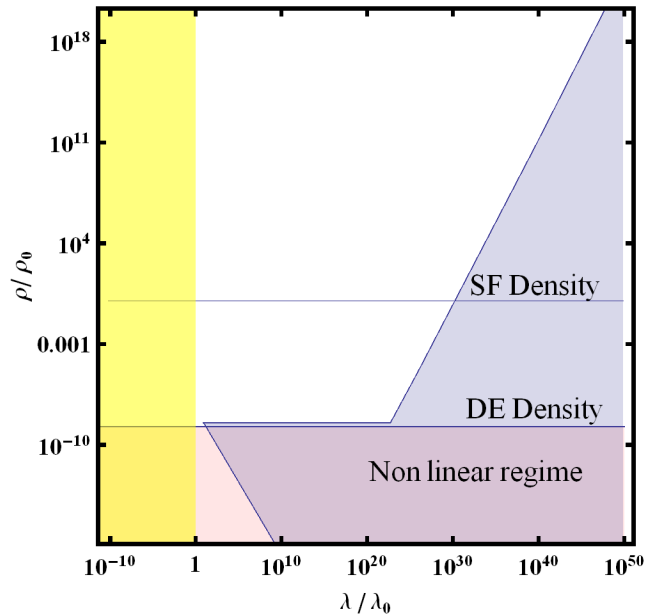}}
\caption {\footnotesize
Analogous to Figure \ref{Validity regime 1} for studied model 3. The limit of the region depicted in yellow shows the proposed value
$\lambda=0.69$ \cite{Starobinsky:2007hu}.}
\label{Validity model 3}
\end{center}
\end{figure}

\section{V. Conclusions}
In this work we have studied the gravitational collapse in $f(R)$ gravity theories.  These theories
provide corrections to the field equations that modify the evolution of gravitational collapse with respect to the
usual General Relativity results. In this context, viable $f(R)$ models must provide similar results for the collapse times
to the values obtained in General Relativity. In addition, collapse times must be much shorter than
the age of the universe and long enough to allow matter cluster.

The analyzed $f(R)$ models present both 
important different quantitative
and 
qualitative behaviors when compared with General Relativity collapses. In fact, all of them show
a collapsing initial epoch with higher contraction than in General Relativity. This result is expected since
$f(R)$ theories modify the gravitational interaction by the addition of a new scalar mediator. It is well-known
that a scalar force is always attractive and can only reduce the time of gravitational collapse. This result is
interesting since observations of structures at high redshift introduce some tension with the standard
$\Lambda$CDM model \cite{clusters}, and the tendency of $f(R)$ models to increase the gravitational attraction
at early times can alleviate this problem.

Although this general behavior is shared by the three models analyzed throughout this
investigation, they present significant differences when the modifications to the General Relativity collapse leave the linear regime. On the one hand, the
$R^2$ model has a modification that increases with the density of the collapse object: $(\Delta t_c/t_{GR}) \propto \sqrt{\rho}$. The opposite
behavior is found for the $R^{-1}$ model, where this modification decreases with density as $(\Delta t_c/t_{GR}) \propto \rho^{-6}$. Finally, a
similar situation is reproduced numerically in the Starobinsky model. The departure from the linear collapse is able to exclude interesting
parameters regions of these models that support late-time
acceleration as seen in Figures 2, 4 and 6.

Another relevant question is related to the physics of stellar objects when analyzed in the $f(R)$ modified
gravity theories frame. Although we cannot use straightforwardly the results of this analysis due to the fundamental role that pressure plays
in the stability of these objects, we may get an idea of the importance of the corrections. Inside these objects, pressure is the
same order of magnitude as density, and it is expected to introduce an important modification
into star evolution and dynamics. 
Therefore, it is enough to take into account the typical value of the density of a neutron star, approximately $10^{-3}\, \text{GeV}^4$
to estimate if correction will be important. Although this value is 35 orders of magnitude larger than the
dust density used above, the results do not change dramatically. A direct extrapolation suggests that we may expect even more negligible modifications for $f(R)$ models that present dark energy
scenarios (models 2 and 3). In models with higher powers of the scalar curvature, the correction to General Relativity will be more important but still negligible (as for model 1).

\vspace{1cm}
{\bf Acknowledgments.} This work has been supported by MICINN (Spain) project number FPA 2008-00592 and Consolider-Ingenio MULTIDARK
CSD2009-00064. AdlCD also acknowledges financial support from NRF and URC research fellowships (South Africa)
and kind hospitality of UCM, Madrid while elaborating part of the manuscript.

\end{document}